\documentclass[11pt,aps,prd,showpacs,showkeys,superscriptaddress,preprintnumbers,amsmath,amssymb,nofootinbib]{revtex4-1}

\usepackage{graphicx}
\usepackage{dcolumn}
\usepackage{bm}
\usepackage{color}
\usepackage{epsfig}
\usepackage{amssymb}
\usepackage{amsmath}
\usepackage{amsfonts}
\usepackage{nicefrac}
\usepackage{physics}

\usepackage[toc,page]{appendix}

\usepackage{natbib}
\usepackage{amsmath,amssymb,enumitem}
\usepackage{fontenc}[T1]
\usepackage[utf8]{inputenc}
\usepackage{multirow}
\usepackage{placeins}
\usepackage{slashed}
\usepackage{xspace}
 
\usepackage[usenames, dvipsnames]{xcolor}
\usepackage{url}
\usepackage[linktocpage=true]{hyperref}
\hypersetup{
   colorlinks   = true,
    linkcolor= blue,
    citecolor= blue,
    urlcolor= blue}
    
\usepackage{amsmath}
\usepackage{siunitx}
\usepackage{textcomp}

\usepackage{subfigure}
\usepackage{pgfplots}
\usepackage{tikz}

\usetikzlibrary{decorations.pathmorphing}
\usetikzlibrary{decorations.markings}
\usepgfplotslibrary{external}
\usepackage{mathtools}

\usepackage{caption}

\newcommand{\be}{\begin{equation}}
\newcommand{\ee}{\end{equation}}
\makeatletter

\def\ak{\@ifstar\@@ak\@ak}
\newcommand{\@ak}[1]{\textcolor{ForestGreen}{[\textbf{AK:} #1]}}
\newcommand{\@@ak}[1]{\textcolor{ForestGreen}{#1}}
\makeatother

\newcommand{\pro}[1]{S_{{#1}}(x)}

\newcommand{\tilpro}[1]{\widetilde{S}_{{#1}}(x)}

\newcommand{\qbarq}{\left\langle\overline{q}q\right\rangle}

\newcommand{\gm}[1]{\gamma_{#1}}

\newcommand{\gf}{\gamma_5}

\newcommand{\xbar}{\overline{x}}

\newcommand{\ali}{\alpha_i}

\newcommand{\alg}{\alpha_g}

\newcommand{\alqb}{\alpha_{\overline{q}}}

\newcommand{\rhonp}[1]{\rho^{(+)}_{#1}(s)}

\newcommand{\rhonm}[1]{\rho^{(-)}_{#1}(s)}

\usepackage{ulem}
\newcommand\rsout{\bgroup\markoverwith
      {\textcolor{red}{\rule[0.5ex]{2pt}{0.4pt}}}\ULon}

\begin{document}

\title{Radiative Decays of the Spin-$\nicefrac{3}{2}$ to Spin-$\nicefrac{1}{2}$ Doubly Heavy Baryons in QCD}

\author{T.~M.~Aliev}
\email[Electronic address:~]{taliev@metu.edu.tr}
\affiliation{Physics Department, Middle East Technical University, 06531 Ankara, Turkey}
\author{A.~Ozpineci}
\email[Electronic address:~]{ozpineci@metu.edu.tr}
\affiliation{Physics Department, Middle East Technical University, 06531 Ankara, Turkey}
\author{E.~Askan}
\email[Electronic address:~]{easkan@ankara.edu.tr}
\affiliation{Physics Department, Middle East Technical University, 06531 Ankara, Turkey}
\affiliation{{Physics Department, Ankara University, 06100 Ankara, Turkey}}

\allowdisplaybreaks[2]

\begin{abstract}
    \vspace{1cm}
    The spin-$\nicefrac{3}{2}$ to spin-$\nicefrac{1}{2}$ doubly heavy baryon transition magnetic dipole $G_M$ and electric quadrupole $G_E$ formfactors are calculated in the framework of light cone sum rules method. Moreover, the decay widths of corresponding radiative transitions are estimated. Obtained results of magnetic dipole moments $G_M$ and decay widths are compared with the results present in the literature.

\end{abstract}


\maketitle


\tikzexternaldisable
\section{Introduction}
\label{chp:intro}

The quark model predicts the existence of baryons containing two heavy quarks. The doubly heavy baryons constitute an important and promising system to study the structure of physics of a system containing two heavy quarks as well as for understanding the strong interaction at the non-perturbative domain.

Doubly heavy baryons lie on the focus of theoretical and experimental investigations. Experimentally, the first observation of doubly heavy baryon $\Xi_{cc}^+$ was reported by SELEX Collaboration \cite{xi+}.
The existence of its iso doublet partner $\Xi_{cc}^{++}$ state was observed by LHCb Collaboration in the $\Lambda_c^+K^-\pi^+\pi^+$ spectrum, with the mass $m_{\Xi_{cc}^{++}}=3621.40\pm0.72\pm0.14$ $MeV$ \cite{xi++}. The mass of $\Xi_{cc}^+$ has been updated in \cite{xi++,aaij2,aaij3,aaij4}. In the upgraded experiments planned at LHCb with large data samples additional states predicted by the quark model and new decay channels of doubly heavy baryons may be observed.

The tremendous experimental developments triggered theoretical studies on the physics of doubly heavy baryons. Mainly the theoretical studies are concentrated on studying the mass, electromagnetic properties, and weak decays of doubly heavy baryons. These studies are critical to understanding the flavor structure and dynamics of these baryons. The mass of doubly charmed baryons within different approaches has been studied in many works such as constituent quark model \cite{consQM1,consQM2,consQM3,consQM4,consQM5,consQM6,consQM8,consQM9}, Regge phenomenology \cite{regge1,regge2}, QCD sum rules method \cite{qcdsr1,qcdsr2,qcdsr3,qcdsr4,AlievOctet,AlievDecuplet} and relativistic quark model\cite{rqm1}. The magnetic moment is another important quantity in studying the inner structure of hadrons. The magnetic moment of doubly charmed baryons within different approaches such as the  quark model, chiral perturbation theory, bag model, effective quark mass scheme, lattice QCD and harmonic oscillator model are calculated in \cite{transition15}, \cite{transition10}, \cite{DhirEMS}, \cite{exBag}, \cite{lattice} and \cite{harmonic} respectively.

The spin-$\nicefrac{3}{2}$ to spin-$\nicefrac{1}{2}$ doubly heavy baryon transition magnetic moments receive special attention as it probes the inner structure as well as possible deformation of heavy baryons.

In the present work, we calculate the transition magnetic moments of doubly heavy baryons with spin-$\nicefrac{3}{2}$ to spin-$\nicefrac{1}{2}$ within the light cone QCD sum rules (LCSR). (More about this method can be found in\cite{lcsr1,lcsr2}.) In this method, the operator product expansion (OPE) is performed over twists of non-local operators.

The work is organized as follows. In \textbf{section II} we derive light cone sum rules for the transition magnetic moments of doubly heavy spin-$\nicefrac{3}{2}$ to spin-$\nicefrac{1}{2}$ transition. In \textbf{section III} we present our numerical results on magnetic moments of considered transitions. In this section, we also present a comparison of our results with predictions of other approaches.
\section{Light Cone Sum Rules for Magnetic Moments of Doubly Heavy Baryons}
\label{chp:sr12}

In this section, we will examine the electromagnetic form factors of $\nicefrac{3}{2}\rightarrow\nicefrac{1}{2}$ transitions between the double heavy baryons with quark content $QQ'q$.This transition in the presence of the electromagnetic field is described by
\begin{align}
    \Pi^{\mu\nu}(p,q) = i\int d^4x \int d^4y e^{ipx+iqy}\mel{0}{\mathcal{T}\left\{\eta_{\nicefrac{1}{2}}(x)j_{el}^\nu(y)\overline{\eta}^\mu_{\nicefrac{3}{2}}(0)\right\}}{0}
    \label{initial_corr}
\end{align}
where $j_{el}^\nu$ is the electromagnetic current given by
\begin{equation}
    j_{el}^\nu = \sum_{q'} e_{q'} \bar q' \gamma^\nu q'
\end{equation}
where $e_{q'}$ is the charge of the quark $q'$ and the sum over is over all quark flavors. Note that although the electromagnetic current contains a sum over all quarks, only the terms proportional to the charges of the quarks appearing in the double heavy baryon contribute to the transition at the order studied in this work. Here $\eta_{\nicefrac{1}{2}}$ and $\eta^\mu_{\nicefrac{3}{2}}$ are the interpolating currents of spin-$\nicefrac{1}{2}$ and spin-$\nicefrac{3}{2}$ baryons respectively.

The $J^P$ quantum numbers of doubly heavy baryons are given in Table-\ref{baryontable}. Note that $\Xi^\prime_{QQ^\prime q}$ is antisymmetric under the exchange of heavy quarks whereas the remaining states are symmetric in the flavor space.

\begin{table}[h!]
    \centering
    \begin{tabular}{|c|c|}
        \hline
        State & J$^\text{P}$  \\
        \hline
        $\Xi_{QQq}$&$\nicefrac{1}{2}^+$\\
        \hline
        $\Xi_{QQq}^\ast$&$\nicefrac{3}{2}^+$\\
        \hline
        $\Xi_{QQ^\prime q}$&$\nicefrac{1}{2}^+$\\
        \hline
        $\Xi_{QQ^\prime q}^\ast$&$\nicefrac{3}{2}^+$\\
        \hline
        $\Xi^\prime_{QQ^\prime q}$&$\nicefrac{1}{2}^+$\\
        \hline
    \end{tabular}
    \caption{$J^P$ quantum numbers of doubly heavy $\Xi$ baryons}
    \label{baryontable}
\end{table}

\begin{equation}
    \begin{aligned}
        \eta^{(S)} &= \dfrac{N}{2}\epsilon^{abc}\sum_{k=1}^2\left[\left(Q^{aT}CA_1^kq^b\right)A_2^kQ^{\prime c}+\left(Q^{\prime aT}CA_1^kq^b\right)A_2^kQ^c\right] \\
        \eta^{(A)} &= \dfrac{1}{\sqrt{6}}\epsilon^{abc}\sum_{k=1}^2\left[2\left(Q^{aT}CA_1^kQ^{\prime b}\right)A_2^kq^c + \left(Q^{aT}CA_1^kq^b\right)A_2^kQ^{\prime c} - \left(Q^{\prime aT}CA_1^kq^b\right)A_2^kQ^{c}\right]
    \end{aligned}
    \label{spin12currents}
\end{equation}
where $a,b,c$ are color indices, $N$ is the normalization constant which is equal to $\sqrt{2}$ when $Q\neq Q^\prime$, $1$ otherwise, $C$ is the charge conjugation matrix, $T$ is the transposition, $A_1^1 = 1$, $A_1^2 = \beta\gamma_5$, $A_2^1 = \gamma_5$ and $A_2^2 = \beta$, and $\beta$ is an arbitrary parameter. Superscripts $(S)$ and $(A)$ mean symmetric and anti-symmetric interpolating currents with respect to the exchange of heavy quarks.

The interpolating current of doubly heavy baryon with spin-$\nicefrac{3}{2}$ can be written as

\begin{equation}
    \eta^{\mu} = \dfrac{N}{\sqrt{3}}\varepsilon^{abc}\left[\left(q^{aT}C\gamma^{\mu}Q^b\right)Q^{\prime c} + \left(q^{aT}C\gamma^{\mu}Q^{\prime b}\right)Q^c + \left(Q^{aT}C\gamma^\mu Q^{\prime b}\right)q^c\right]
    \label{spin32current}
\end{equation}
where again $a,b,c$ are color indices, $N$ is the normalization constant which is equal to $\sqrt{2}$ when $Q\neq Q^\prime$, $1$ otherwise, $C$ is the charge conjugation operator. Note that this interpolating current is symmetric under the exchange of any two quark fields.

By introducing the plane wave electromagnetic background field

\begin{equation}
    F_{\mu\nu} = i\left(\varepsilon_\nu^{(\lambda)}q_\mu-\varepsilon_\mu^{(\lambda)}q_\nu\right)e^{iqx}
    \label{backgroundfield}
\end{equation}
and multiplying the correlation function (\ref{initial_corr}) by $\varepsilon_\nu^{(\lambda)}$, it can be rewritten in the following form
\begin{equation}
    \Pi^{\mu\nu}(p,s) \varepsilon_\nu^{(\lambda)} = i\int d^4x e^{ipx} \mel{0}{\mathcal{T}\left\{\eta_{\nicefrac{1}{2}}(x)\overline{\eta}^\mu_{\nicefrac{3}{2}}(0)\right\}}{0}_F
    \label{spin12bfcorr}
\end{equation}

In Eq-(\ref{spin12bfcorr}) the subscript $F$ indicates that all vacuum expectation values are evaluated in the presence of the background field. 
It should be noted that the correlation function given in Eq-(\ref{initial_corr}) can be obtained from Eq-(\ref{spin12bfcorr}) by expanding latter in powers of the background field and taking into account only terms linear in $F_{\mu\nu}$, which corresponds to the single photon emission. (More about the background field method can be found in \cite{PhotonDA} and \cite{ExternalField}.)

Now let us derive the sum rules for the $\nicefrac{3}{2}\longrightarrow\nicefrac{1}{2}$ transition form factors.  
According to the sum rules philosophy, the correlation function Eq-(\ref{initial_corr}) has to be calculated in two different kinematic domains. 
The hadronic representation of the correlation function can be calculated by inserting a full set of hadrons carrying the same quantum numbers as the interpolating currents $\eta_{\nicefrac{1}{2}}$ and $\eta^\mu_{\nicefrac{3}{2}}$ and 
then separating the contributions of ground states. On the other hand, the correlation function can be calculated in deep Euclidean region, where $p^2\ll0$ and $(p+q)^2\ll0$ with the help of operator product expansion (OPE). 
The OPE is performed in terms of photon distribution amplitudes(DAs) of increasing twist. 

At the hadronic level, the correlation function Eq-(\ref{spin12bfcorr}) can be written as
\begin{equation}
    \Pi^{\mu\nu}\epsilon_\nu^{(\lambda)} = \dfrac{\epsilon_\nu^{(\lambda)}\mel{0}{\eta_{\nicefrac{1}{2}}}{B_2(p,s)}\mel{B_2(p,s)}{j_{el}^\nu}{B_1(p+q,s^\prime}\mel{B_1(p+q,s^\prime)}{\overline{\eta}^\mu_{\nicefrac{3}{2}}}{0}}{\left(p^2-m_2^2\right)\left((p+q)^2-m_1^2\right)}
    \label{corr_with_inserted_states}
\end{equation}
where a sum over hadrons and the continuum is not shown explicitly.
The matrix elements in eq-(\ref{corr_with_inserted_states}) are defined as
\begin{equation}
    \begin{aligned}
        \mel{0}{\eta_{\nicefrac{1}{2}}}{B_{\nicefrac{1}{2}}(p,s)} &= \lambda_2 u(p,s)\\
        \mel{B_{\nicefrac{3}{2}}(p+q,s)}{\overline{\eta}^\mu_{\nicefrac{3}{2}}}{0} &= \lambda_1 \overline{u}^\mu(p+q,s)
    \end{aligned}
    \label{mels}
\end{equation}
where $u$ is a Dirac spinor, $u^\mu$ is a Rarita-Schwinger spinor and $\lambda_{1\left(2\right)}$ is the residue of the spin-$\nicefrac{3}{2}\left(\nicefrac{1}{2}\right)$ doubly heavy baryon.

The matrix element of the  electromagnetic current sandwiched between spin-$\nicefrac{1}{2}$ and spin-$\nicefrac{3}{2}$ states is parametrized in terms of three form factors in the following way \cite{JONESmatrixel,KORNERmatrixel}:
\begin{equation}
    \begin{aligned}
        \mel{B_{\nicefrac{1}{2}}(p,s)}{j_{el}^\nu}{B_{\nicefrac{3}{2}}(p+q,s)} &=\overline{u}(p,s)\Bigg[G_1\left(-g^{\alpha\nu}\slashed{q}+q^{\alpha}\gamma^{\nu}\right)\\
        &+G_2\left(-g^{\alpha\nu}\left(p+\dfrac{q}{2}\right)q + q^\alpha\left(p+\dfrac{q}{2}\right)^\nu\right)\\
        &+G_3\left(q^\alpha q^\nu - q^2g^{\alpha\nu}\right)\Bigg]\gf u_\alpha(p+q,s)
    \end{aligned}
    \label{el_mel}
\end{equation}
The last term which is proportional to $ G_3$ does not contribute to the emission of a real photon. In the problem studied in this work, the emitted photon is a real photon. For this reason, only the values of $G_1$ and $G_2$ at $q^2=0$ are needed.

Magnetic dipole ($G_E$) and electric quadrupole ($G_E$) moments, which are more directly accessible in experiments, are defined through the formfactors $G_1$ and $G_2$ as
\cite{JONESmatrixel,KORNERmatrixel}:
\begin{equation}
    \begin{aligned}
        G_M &= \Bigg[\dfrac{3m_1+m_2}{m_1}G_1(0) + \left(m_1-m_2\right)G_2(0)\Bigg]\dfrac{m_2}{3}\\
        G_E &= \left(m_1-m_2\right)\Bigg[\dfrac{G_1(0)}{m_1}+G_2(0)\Bigg]\dfrac{m_2}{3}
    \end{aligned}
    \label{GMandGE}
\end{equation}
where $m_1$ and $m_2$ are masses of spin-$\nicefrac{3}{2}$ and spin-$\nicefrac{1}{2}$ doubly heavy baryons, respectively.
Performing summation over spins for the Dirac and Rarita-Schwinger spinors via
\begin{equation}
    \begin{aligned}
        \sum u(p,s)\overline{u}(p,s) &= \left(\slashed{p}+m\right)\\
        \sum u_\alpha(p,s)\overline{u}_\beta(p,s) &= \left(\slashed{p}+m\right)\left[-g_{\alpha\beta}+\dfrac{1}{3}\gamma_\alpha\gamma_\beta-\dfrac{2p_\alpha p_\beta}{3m^2}-\dfrac{p_\alpha\gamma_\beta-p_\beta\gamma_\alpha}{3m}\right]
    \end{aligned}
    \label{spin_summations}
\end{equation}
the phenomenological representation of the correlation function can be obtained. In order to obtain the sum rules for transition form factors, two issues need to be addressed:
\begin{enumerate}
    \item Besides the spin-$\nicefrac{3}{2}$ baryon, the current $\eta_\mu$ can also create a negative parity baryon with spin-$\nicefrac{1}{2}$ from the vacuum. This negative parity baryon also contributed to the correlation function.
    \item Not all Lorentz structures are independent of each other.
\end{enumerate}
The first problem can be solved in the following way. The matrix element of the current $\eta_\mu$ between vacuum and a spin-$\nicefrac{1}{2}$ baryon can be written as
\begin{equation}
    \mel{B_{\nicefrac{1}{2}^-}(p+q,s)}{\eta_\mu}{0} = A \bar u(p+q) \left[(p+q)_\mu-\dfrac{m}{4}\gamma_\mu\right]
\end{equation}
where the fact that $\eta_\mu \gamma^\mu=0$ is used. Hence, it is seen that the contributions of the spin-$\nicefrac{1}{2}$ baryons are either $\propto (p+q)_\mu$ or have $\gamma_\mu$ at the right. Other structures do not receive any contribution from these spin-$\nicefrac{1}{2}$ baryons, and only receive contribution from spin-$\nicefrac{3}{2}$ baryons.
In order to solve the second issue and obtain independent Lorentz structures (following \cite{AlievRadDec}) the Dirac matrices are ordered as $\slashed{\varepsilon}\slashed{q}\slashed{p}\gamma_\mu$. With this ordering, the hadronic representation of the correlation function becomes:
    \begin{align}
        \nonumber\Pi_{\mu\nu}(p,q)\varepsilon^\nu =& e \lambda_2 \lambda_1 \frac{1}{p^{2}-m_2^{2}} \frac{1}{(p+q)^{2}-m_1^{2}}\Bigg\{\\
        \nonumber&{\left[\varepsilon_{\mu}(p q)-(\varepsilon p) q_{\mu}\right]\left\{-2 G_{1} m_1-G_{2} m_1 m_2+G_{2}(p+q)^{2}\right.}\\
        \nonumber+&\left.\left[2 G_{1}+G_{2}\left(m_2-m_1\right)\right] \slashed{p}+m_2 G_{2} \slashed{q}-G_{2} \slashed{q} \slashed{p}\right\} \gamma_{5}\\
        +&\left[q_{\mu} \slashed{\varepsilon}-\varepsilon_{\mu} \slashed{q}\right]\left\{G_{1}\left(p^{2}+m_1 m_2\right)-G_{1}\left(m_1+m_2\right) \slashed{p}\right\} \gamma_{5}\\
        \nonumber+&2 G_{1}[\slashed{\varepsilon}(p q)-\slashed{q}(\varepsilon p)] q_{\mu} \gamma_{5}\\
        \nonumber-&G_{1} \slashed{\varepsilon} q\{m+\slashed{p}\} q_{\mu} \gamma_{5}\\
        \nonumber+&\text { other structures with } \gamma^{\mu} \text { at the end}\\
        \nonumber+&\text{structures which are proportional to }(p+q)^{\mu}\Bigg\}
    \end{align}
Now let us turn our attention to the calculation of the correlation function from the QCD side. After applying Wick theorem the correlation function given in Eq.-(\ref{spin12bfcorr}) becomes:

    \begin{align}
        \nonumber\Pi_\mu^S(p) &= i\int d^4xe^{ipx}\dfrac{6}{\sqrt{6}}\bra{0}\left\{\pro{Q'}\gamma_5\Tr\Big[\gamma_\mu\pro{Q}\tilpro{q}\Big]\right.\\
        \nonumber&-\pro{Q}\tilpro{q}\gamma_\mu\pro{Q'}\gamma_5-\pro{q}\tilpro{Q}\gamma_\mu\pro{Q'}\gamma_5\\
        \nonumber&-\pro{Q'}\tilpro{q}\gamma_\mu\pro{Q}\gamma_5+\pro{Q}\gamma_5\Tr\Big[\tilpro{q}\gamma_\mu\pro{Q'}\Big]\\
        \nonumber&+\pro{q}\tilpro{Q'}\gamma_\mu\pro{Q}\gamma_5+\beta\pro{Q'}\Tr\Big[\gamma_\mu\pro{Q}\gamma_5\tilpro{q}\Big]\\
        \nonumber&-\beta\pro{Q}\gamma_5\tilpro{q}\gamma_\mu\pro{Q'}-\beta\pro{q}\gamma_5\tilpro{Q}\gamma_\mu\pro{Q'}\\
        \nonumber&-\beta\pro{Q'}\gamma_5\tilpro{q}\gamma_\mu\pro{Q}+\beta\pro{Q}\Tr\Big[\tilpro{q}\gamma_\mu\pro{Q'}\gamma_5\Big]\\
        \nonumber&\left.+\beta\pro{q}\gamma_5\tilpro{Q'}\gamma_\mu\pro{Q}\right\}\ket{0}_F\\
        \label{transition_after_wick}\Pi_\mu^A(p) &= i\int d^4xe^{ipx}\dfrac{6}{3\sqrt{2}}\bra{0}\left\{2\pro{Q'}\tilpro{Q}\gm{\mu}\pro{q}\gf-\beta\pro{q}\gf\tilpro{Q'}\gm{\mu}\pro{Q}\right.\\
        \nonumber&+2\pro{Q}\tilpro{Q'}\gm{\mu}\pro{q}\gf +2\pro{q}\gf\Tr\Big[\tilpro{Q}\gm{\mu}\pro{Q'}\Big]\\
        \nonumber&-\pro{Q'}\gf\Tr\Big[\pro{Q}\gm{\mu}\pro{q}\Big] -\pro{Q}\tilpro{q}\gm{\mu}\pro{Q'}\gf\\
        \nonumber&-\pro{q}\tilpro{Q}\gm{\mu}\pro{Q'}\gf +\pro{Q'}\tilpro{q}\gm{\mu}\pro{Q}\gf\\
        \nonumber&+\pro{Q}\gf\Tr\Big[\tilpro{Q'}\gm{\mu}\pro{q}\Big] -\pro{q}\tilpro{Q'}\gm{\mu}\pro{Q}\gf\\
        \nonumber&+2\beta\pro{Q'}\gf\tilpro{Q}\gm{\mu}\pro{q} -2\beta\pro{Q}\gf\tilpro{Q'}\gm{\mu}\pro{q}\\
        \nonumber&+2\beta\pro{q}\Tr\Big[\gm{\mu}\pro{Q'}\gf\tilpro{Q}\Big] -\beta\pro{Q'}\Tr\Big[\gf\tilpro{q}\gm{\mu}\pro{Q}\Big]\\
        \nonumber&-\beta\pro{Q}\gf\tilpro{q}\gm{\mu}\pro{Q'} -\beta\pro{q}\gf\tilpro{Q}\gm{\mu}\pro{Q'}\\
        \nonumber&\left.+\beta\pro{Q'}\gf\tilpro{q}\gm{\mu}\pro{Q} +\beta\pro{Q}\Tr\Big[\gf\tilpro{Q'}\gm{\mu}\pro{q}\Big]\right\}\ket{0}_F
    \end{align}
where the superscripts $S(A)$ denotes when symmetric(anti-symmetric) interpolating current has been used. In Eq-(\ref{transition_after_wick}), $S_Q$ and $S_q$ are heavy and light quark propagators. Their expressions in the presence of gluonic and electromagnetic background fields are
\begin{equation}
    \begin{aligned}
        S_q(x) &= \dfrac{i\slashed{x}}{2\pi^2x^4}-\dfrac{i}{16\pi^2x^2}\int_0^1du\Bigg\{\overline{u}\slashed{x}\sigma_{\alpha\beta}+u\sigma_{\alpha\beta}\slashed{x}\Bigg\}\Bigg\{gG^{\alpha\beta}(ux)+e_qF^{\alpha\beta}(ux)\Bigg\}\\
        S_Q(x) &= \dfrac{m_Q^2}{4\pi^2}\left[\dfrac{K_1\left(m_Q\sqrt{-x^2}\right)}{\sqrt{-x^2}}+\dfrac{i\slashed{x}}{\left(\sqrt{-x^2}\right)^2}K_2\left(m_Q\sqrt{-x^2}\right)\right]\\
        &-\dfrac{im_Q}{16\pi^2}\int_0^1 du\Bigg\{gG^{\alpha\beta}(ux) + e_QF^{\alpha\beta}(ux)\Bigg\}\\
        &\Bigg\{\sigma_{\alpha\beta}K_0\left(m_Q\sqrt{-x^2}\right) + \left(\left(\overline{u}\slashed{x}\sigma_{\alpha\beta}+u\sigma_{\alpha\beta}\slashed{x}\right)\dfrac{K_1\left(m_Q\sqrt{-x^2}\right)}{\sqrt{-x^2}}\right)\Bigg\}
    \end{aligned}
    \label{propagators}
\end{equation}
Where $K_i$ are the modified Bessel functions of the second kind, $G_{\mu\nu}$ and $F_{\mu\nu}$ are the background gluon and electromagnetic field strength tensors respectively.

Calculation of the correlation function in the external field involves perturbative and non-perturbative parts, i.e. the photon interacts with the quarks perturbatively or non-perturbatively. To calculate the perturbative part, when the propagators in Eq.-(\ref{propagators}) is placed in Eq.-(\ref{transition_after_wick}), the terms linear in $F_{\mu \nu}$ are selected. For the calculation of the non-perturbative part, it is enough to replace the light quark propagator that emits the photon with
\begin{equation}
    S_q \longrightarrow -\dfrac{1}{4}\Gamma_j\overline{q}\Gamma_j q
\end{equation}
where $\Gamma_j = \left\{\mathbf{1},\gamma_\alpha, i\gamma_5\gamma_\alpha, \gamma_5, \dfrac{1}{\sqrt{2}}\sigma_{\alpha\beta}\right\}$. In this case, there are two and three particle matrix elements: $\mel{0}{\overline{q}\Gamma_j q}{0}_F$, $\mel{0}{\overline{q}\Gamma G_{\mu\nu}q}{0}_F$ and $\mel{0}{\overline{q}\Gamma F_{\mu\nu}q}{0}_F$ which are parametrized in terms of photon DAs and describes the interaction of photons with quark fields at large distance.

We see that the hadronic and QCD sides of the correlation function contain many structures. Among all structures, the structures $\slashed{\varepsilon}\slashed{p}\gf q_\mu$ and $\slashed{q}\slashed{p}\gf\left(\varepsilon p\right)q_\mu$ exhibit the best convergence, for this reason, we choose them for determination of $G_1(0)$ and $G_2(0)$ respectively. equating the coefficients of the aforementioned structures on both sides, we get the sum rules for $G_1(0)$ and $G_2(0)$. In the final step, we perform Borel transformation over variables $-p^2$ and $-(p+q)^2$ in order to suppress the higher states and continuum contributions and enhance ground states to obtain
\begin{equation}
    \begin{aligned}
        \left(m_2+m_1\right)G_1(0)\lambda_1\lambda_2e^{-\nicefrac{m_2^2}{M_1^2}}e^{-\nicefrac{m_1^2}{M_2^2}}  +\cdots &= \int ds_1 \int ds_2 e^{-\nicefrac{s_1}{M_1^2}-\nicefrac{s_2}{M_2^2}}\rho_1(s_1,s_2)\\
        G_2(0)\lambda_1\lambda_2e^{-\nicefrac{m_2^2}{M_1^2}}e^{-\nicefrac{m_1^2}{M_2^2}} +\cdots&= \int ds_1 \int ds_2 e^{-\nicefrac{s_1}{M_1^2}-\nicefrac{s_2}{M_2^2}}\rho_2(s_1,s_2)\\
    \end{aligned}
    \label{correlated}
\end{equation}
where $\lambda_1$ and $\lambda_2$ are the residues of the spin-$\nicefrac{3}{2}$ and spin-$\nicefrac{1}{2}$ baryons respectively, and $\cdots$ denote the contribution from higher states and the continuum. To obtain the sum rules for the form factors $G_1$ and $G_2$, the contributions of the higher states and the continuum are subtracted using quark hadron duality: 
\begin{equation}
    \rho(s_1,s_2) \simeq \rho^{OPE}(s_1,s_2) ~~\mbox{if}~~ (s_1,s_2) \not\in {\cal D}
\end{equation}
where $\cal D$ is a domain in the $(s_1,s_2)$ plane.
Typically, the domain $\cal D$ is a rectangular region defined by $s_1<s_{10}$ and $s_2<s_{20}$ for some constants $s_{10}$ and $s_{20}$, or a triangular region. In this work, for its simplicity, continuum subtraction is carried out
by choosing $\cal D$ as the region defined as $s \equiv s_1 u_0 + s_2 \bar u_0 < s_0$ where $u_0\equiv \frac{M_2^2}{M_1^2+M_2^2}$ and $\bar u_0 = 1 - u_0$. Introducing a second variable $u=\frac{s_1 u_0}{s}$, the integral in the $(s_1,s_2)$ plane can be written as:
\begin{equation}
    \int ds_1 \int ds_2 e^{-s_1/M_1^2-s_2/M_2^2} \rho(s_1,s_2) = \int ds e^{-\frac{s}{M^2}} \rho(s)
\end{equation}
where 
\begin{equation}
    \rho(s) = \frac{s}{u_0 \bar u_0} \int_0^1 du \rho\left( s \frac{u}{u_0}, s \frac{\bar u}{\bar u_0} \right) 
\end{equation}
In the problem under study, the masses of the initial and final state baryons are very close to each other, hence we can set $M_1^2 = M_2^2 = 2M^2$ leading to $u_0=\nicefrac{1}{2}$. 
Using quark hadron duality, sum rules for the form factors take the form:
\begin{equation}
    \begin{aligned}
        \left(m_1+m_2\right)G_1(0)\lambda_1\lambda_2e^{-\nicefrac{m^2}{M^2}} &= \int_0^{s_0} ds e^{-\nicefrac{s}{M^2}}\rho_1(s) \\
        G_2(0)\lambda_1\lambda_2e^{-\nicefrac{m^2}{M^2}} &= \int_0^{s_0} ds e^{-\nicefrac{s}{M^2}}\rho_2(s)
    \end{aligned}
    \label{correlated_with_s}
\end{equation}
where $m^2=(m_1^2+m_2^2)/2$ and the expressions of spectral densities $\rho_1(s)$ and $\rho_2(s)$ are presented in Appendix-\ref{appndxb}. Note that although the limits of the $s$ integral is written from $s=0$ upto $s=s_0$, $\rho_i(s)=0$ ($i=1$ or $2$) for $s < (m_Q+m_{Q'})^2$.
From Eq-(\ref{correlated_with_s}) it follows that, for determination of $G_1(0)$ and $G_2(0)$ the residues $\lambda_1$ and $\lambda_2$ are needed. Spin-$\nicefrac{1}{2}$ residues are calculated in \cite{AlievOctet}, while the values for the spin-$\nicefrac{3}{2}$ residues are calculated in \cite{AlievDecuplet}.

\section{Numerical Analysis}
\label{chp:rslts12}

This section is devoted on the analysis of the sum rules for the transition formfactors $G_1(0)$ and $G_2(0)$. For the values, the input parameters appearing in the sum rules are

\begin{table}[!ht]
    \centering
    \begin{tabular}{crl}
        $m_c\left(\overline{m}_c\right)$ &=& $\left(1.275 \pm 0.025\right) \text{GeV}$ \cite{PDG} \\
        $m_b\left(\overline{m}_b\right)$ &=& $\left(4.18 \pm 0.03\right) \text{GeV}$ \cite{PDG} \\
        $f_{3\gamma}$ &=& $-0.0039 \text{GeV}^2$ \cite{PhotonDA} \\
        $\chi$ &=& $\left(3.15\pm0.10\right) \text{GeV}^2$ \cite{PhotonDA} \\
        $\qbarq$ &=& $\left(-0.24 \pm 0.001\right)^3 \text{GeV}^3$ \cite{ioffeCondensate} \\
        $m_0^2$ &=& $\left(0.8 \pm 0.2\right) \text{GeV}^2$ \cite{SVZ}
    \end{tabular}
    \caption*{}
    \label{inputParams}
\end{table}
For the heavy quark masses, we have used their $\overline{\text{MS}}$ Scheme values. The masses of the spin-$\nicefrac{3}{2}$ doubly heavy baryons and spin-$\nicefrac{1}{2}$ doubly heavy baryons are calculated in \cite{lattice_masses,AlievOctet,AlievDecuplet}, and the mass of $\Xi_{cc}$ baryon is experimentally observed.

\begin{table}[ht]
    \centering
    \begin{tabular}{|c|c|c|c|c|}
         \hline
         Baryon & Lattice\cite{lattice_masses} &  QCDSR\cite{AlievOctet} & QCDSR\cite{AlievDecuplet} & Exp\cite{xi+,xi++}\\
         \hline
         $\Xi_{bc}$ & 6.943 GeV & $6.72 GeV$ & - & -  \\
         \hline
         $\Xi_{bc}^{\prime}$ & 6.959 GeV & $6.79 GeV$ & - & - \\
         \hline
         $\Xi_{bb}$ & 10.143 GeV & $9.96 GeV$ & - & - \\
         \hline
         $\Xi_{cc}$ & 3.610 GeV& $3.72 GeV$ & - & $3.52 GeV$  \\
         \hline
         $\Xi^{\ast}_{bc}$ & 6.985 GeV& - & $7.25 GeV$& -  \\
         \hline
         $\Xi^{\ast}_{cc}$ & 3.692 GeV& - &  $3.69  GeV$& - \\
         \hline
         $\Xi^{\ast}_{bb}$ & 10.178 GeV & - & $10.4  GeV$& -  \\
         \hline
    \end{tabular}
    \caption{Baryon masses}
    \label{masstable}
\end{table}
We will see that the radiative decay widths of spin-$\nicefrac{3}{2}$ to spin-$\nicefrac{1}{2}$ baryons are proportional to the cube of the mass difference of the initial and final baryons, $\left(\Delta m\right) = \left(m_{\nicefrac{3}{2}}-m_{\nicefrac{1}{2}}\right)$. Therefore, the decay widths are very sensitive to the mass difference of doubly heavy baryons. For this reason, in the next discussion, we will use the mass of doubly heavy baryons obtained from lattice calculations which contain a minimal error.
The photon distribution amplitudes are the main non-perturbative inputs of light cone sum rules. The expression of DAs and the values of the parameters entering in the expressions of DAs are given in \cite{PhotonDA}. Now, let us perform the numerical analysis of the relevant formfactors.

The sum rules for the transition formfactors $G_1(0)$ and $G_2(0)$ involve three auxiliary parameters: parameter $\beta$ appearing in the expression of the interpolating current for spin-$\nicefrac{1}{2}$ doubly heavy baryons, the continuum threshold $s_0$ and Borel mass parameter $M^2$. The working region of $s_0$ is determined from the analysis of two-point sum rules which are carried out in \cite{AlievOctet,AlievDecuplet} and given in Table-\ref{parameter_table}

The working region of $M^2$ is determined from two requirements. From one side $M^2$ must be large enough to guarantee the dominance of leading twist over higher twist contributions, and from the other side, it should be small enough in order to ensure the suppression of the higher states and continuum contributions. The working regions of Borel mass parameters are also presented in Table-\ref{parameter_table}.

As an example, in Figures-\ref{xibcpgms}(a) and \ref{xibcpges}(a) we present the dependencies of $G_M(0)$ and $G_E(0)$ on $M^2$ at fixed values of $s_0$ at $\beta$. From these figures, we see that the formfactors $G_M$ and $G_E$ exhibit excellent stability with respect to the variation of $M^2$ in its working region.

\tikzexternalenable
\label{chp:appndxd}

\pgfplotsset{width=10cm,compat=1.18}


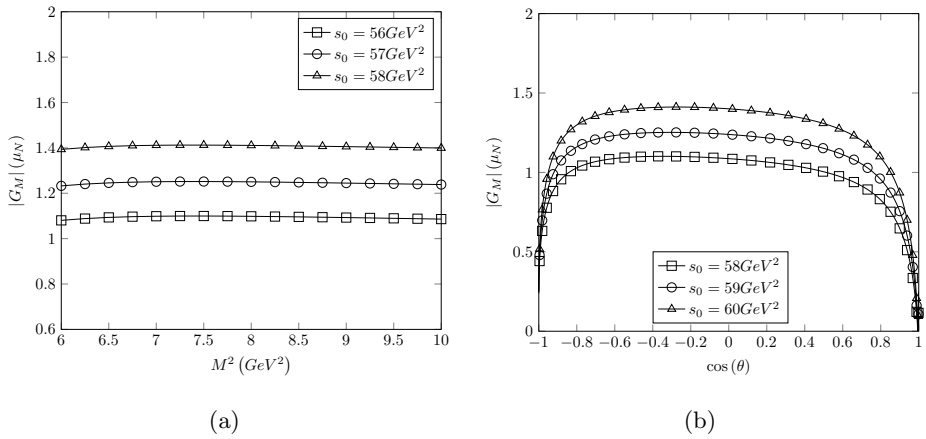
\begin{figure}[h!]
    \centering
    \subfigure[]{
    \begin{tikzpicture}[scale=0.6]
		\begin{axis}[
			xlabel={$M^2 \left(GeV^2\right)$},
			ylabel={$|G_M| \left(\mu_N\right)$},
			enlargelimits=false,
			ymin=0.6,
			ymax=2,
			]
			\addplot [
			color=black,
			mark=square,
			mark size=2.9pt]
			table[x={Msq}, y={gmlow}]{dataforfigs/xibcpgs.dat};
			\addlegendentry{$s_0 = 56 GeV^2$}
			\addplot [
			color=black,
			mark=o,
			mark size=2.9pt]
			table[x={Msq}, y={gmmid}]{dataforfigs/xibcpgs.dat};
			\addlegendentry{$s_0 = 57 GeV^2$}
			\addplot [
			color=black,
			mark=triangle,
			mark size=2.9pt]
			table[x={Msq}, y={gmhigh}]{dataforfigs/xibcpgs.dat};
			\addlegendentry{$s_0 = 58 GeV^2$}
		\end{axis}
	\end{tikzpicture}
    }
    \subfigure[]{
    \begin{tikzpicture}[scale=0.6]
    \pgfplotsset{
            every axis legend/.append style={
                at={(0.3,0.025)},
                anchor=south west,
            },
        }
		\begin{axis}[
			xlabel={$\cos\left(\theta\right)$},
			ylabel={$|G_M| \left(\mu_N\right)$},
			enlargelimits=false,
			ymin=0,
			ymax=2,
			]
			\addplot [
			color=black,
			mark=square,
			mark repeat=10,
			mark phase=10,
			mark size=2.9pt]
			table[x={cos}, y={gm58}]{dataforfigs/xibcpcosthetas.dat};
			\addlegendentry{$s_0 = 58 GeV^2$}
			\addplot [
			color=black,
			mark=o,
			mark repeat=10,
			mark phase=10,
			mark size=2.9pt]
			table[x={cos}, y={gm59}]{dataforfigs/xibcpcosthetas.dat};
			\addlegendentry{$s_0 = 59 GeV^2$}
			\addplot [
			color=black,
			mark=triangle,
			mark repeat=10,
			mark phase=10,
			mark size=2.9pt]
			table[x={cos}, y={gm60}]{dataforfigs/xibcpcosthetas.dat};
			\addlegendentry{$s_0 = 60 GeV^2$}
		\end{axis}
	\end{tikzpicture}
    }
    \caption[$G_M$ for $\Xi^{\ast}+_{bc}\rightarrow\Xi^{+}_{bc}\gamma$]{\textbf{a)} $M^2$ dependence of the $G_M$ for $\Xi^{\ast+}_{bc}\rightarrow\Xi^{+}_{bc}\gamma$ decay with different values of $s_0$ at $\beta=5$, \textbf{b)} $\cos\left(\theta\right)$ dependence of $G_M$ for $\Xi^{\ast+}_{bc}\rightarrow\Xi^{+}_{bc}\gamma$ with different values of $s_0$ at $M^2=8$ $GeV^2$ where $\beta=\tan\left(\theta\right)$}
    \label{xibcpgms}
\end{figure}

\begin{figure}[h!]
    \centering
    \subfigure[]{
    \begin{tikzpicture}[scale=0.6]
		\begin{axis}[
			xlabel={$M^2 \left(GeV^2\right)$},
			ylabel={$|G_E| \left(10^{-3}\mu_N\right)$},
			enlargelimits=false,
			ymin=0,
			ymax=5,
			]
			\addplot [
			color=black,
			mark=square,
			mark size=2.9pt]
			table[x={Msq}, y={gelow}]{dataforfigs/xibcpges.dat};
			\addlegendentry{$s_0 = 58 GeV^2$}
			\addplot [
			color=black,
			mark=o,
			mark size=2.9pt]
			table[x={Msq}, y={gemid}]{dataforfigs/xibcpges.dat};
			\addlegendentry{$s_0 = 59 GeV^2$}
			\addplot [
			color=black,
			mark=triangle,
			mark size=2.9pt]
			table[x={Msq}, y={gehigh}]{dataforfigs/xibcpges.dat};
			\addlegendentry{$s_0 = 60 GeV^2$}
		\end{axis}
	\end{tikzpicture}  
    }
	\subfigure[]{
    \begin{tikzpicture}[scale=0.6]
		\begin{axis}[
			xlabel={$\cos\left(\theta\right)$},
			ylabel={$|G_E| \left(10^3\mu_N\right)$},
			enlargelimits=false,
			ymin=0,
			ymax=5,
			]
			\addplot [
			color=black,
			mark=square,
			mark repeat=10,
			mark phase=10,
			mark size=2.9pt]
			table[x={cos}, y={ge58}]{dataforfigs/xibcpcosthetas.dat};
			\addlegendentry{$s_0 = 58 GeV^2$}
			\addplot [
			color=black,
			mark=o,
			mark repeat=10,
			mark phase=10,
			mark size=2.9pt]
			table[x={cos}, y={ge59}]{dataforfigs/xibcpcosthetas.dat};
			\addlegendentry{$s_0 = 59 GeV^2$}
			\addplot [
			color=black,
			mark=triangle,
			mark repeat=10,
			mark phase=10,
			mark size=2.9pt]
			table[x={cos}, y={ge60}]{dataforfigs/xibcpcosthetas.dat};
			\addlegendentry{$s_0 = 60 GeV^2$}
		\end{axis}
	\end{tikzpicture}
    }
    \caption[$G_E$ for $\Xi^{\ast+}_{bc}\rightarrow\Xi^{+}_{bc}\gamma$]{Same as \ref{xibcpgms} but for $G_E$}
    \label{xibcpges}
\end{figure}
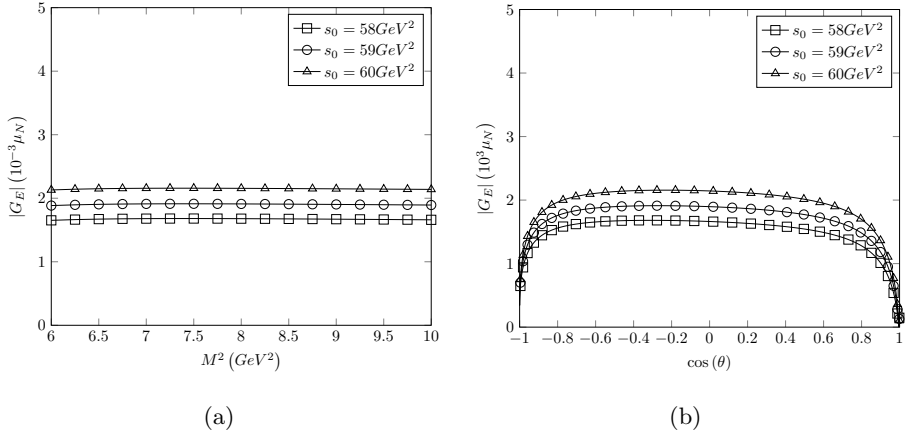

Finally in order to find the working region of $\beta$, we present the dependence of $G_M(0)$ and $G_E(0)$ for the $\Xi_{bc}^{\ast+}\rightarrow\Xi_{bc}^{+}$ transition on $\cos(\theta)$ where $\beta=\tan(\theta)$ in Figures-\ref{xibcpgms}(b) and \ref{xibcpges}(b). From these figures, we see that the results on the formfactors show very good stability when $-0.4<\cos(\theta)<0.4$. Performing similar calculations for the formfactors of other transitions we find that in this region of $\cos(\theta)$ the formfactors are rather stable. Therefore the working region of $\cos\theta$ is $\left(-0.4,0.4\right)$

\begin{table}[ht]
    \centering
    \begin{tabular}{|c|c|c|c|}
        \hline
         Transition & $M^2$ & $\cos(\theta)$ & $s_0$  \\
         \hline
         $\Xi_{bc}^+\rightarrow\Xi_{bc}^{\ast+}$ & $6-9~GeV^2$ & $(-0.4,0.4)$ & $57\pm1 ~GeV^2$\\
         \hline
         $\Xi_{bc}^{\prime+}\rightarrow\Xi_{bc}^{\ast+}$& $6-9~GeV^2$ & $(-0.4,0.4)$ & $57\pm1~ GeV^2$\\
         \hline
         $\Xi_{bc}^0\rightarrow\Xi_{bc}^{\ast0}$& $6-9~GeV^2$ & $(-0.4,0.4)$ & $57\pm1~ GeV^2$\\
         \hline
         $\Xi_{bc}^{\prime0}\rightarrow\Xi_{bc}^{\ast0}$& $6-9 ~GeV^2$ & $(-0.4,0.4)$ & $57\pm1~ GeV^2$\\
         \hline
         $\Xi_{cc}^{++}\rightarrow\Xi_{cc}^{\ast++}$& $3-6 ~GeV^2$ & $(-0.4,0.4)$ & $20\pm1~ GeV^2$\\
         \hline
         $\Xi_{cc}^+\rightarrow\Xi_{cc}^{\ast+}$& $3-6 ~GeV^2$ & $(-0.4,0.4)$ & $20\pm1 ~GeV^2$\\
         \hline
         $\Xi_{bb}^0\rightarrow\Xi_{bb}^{\ast0}$& $9-12 ~GeV^2$ & $(-0.4,0.4)$ & $121\pm2 ~GeV^2$\\
         \hline
         $\Xi_{bb}^-\rightarrow\Xi_{bb}^{\ast-}$& $9-12 ~GeV^2$ & $(-0.4,0.4)$ & $121\pm2 ~GeV^2$\\
        \hline
    \end{tabular}
    \caption{Working Regions for $M^2$, $\beta$ and $s_0$}
    \label{parameter_table}
\end{table}
We performed the numerical analysis by using the working regions of $\cos\theta$, $M^2$ and $s_0$, and our final findings of the moments $G_M(0)$ and $G_E(0)$ are collected in Table-\ref{magnetic_results} and Table-\ref{electric_results}. Note that, since the signs of the residues are undetermined, we present the absolute values of our results.
\begin{table}[!ht]
    \centering
    \begin{tabular}{|c|c|c|c|c|c|c|c|}
        \hline
         Transition & This Work & QM\cite{transition15} & $\chi$ PT\cite{transition15} & Bag Model\cite{transition10} & EMS\cite{DhirEMS} & exBag\cite{exBag} & Lattice\cite{lattice} \\
         \hline
         $\Xi_{bc}^+\rightarrow\Xi_{bc}^{\ast+}$& $1.25\pm0.156$ & $-1.61$ & -2.56 & 0.695 &  1.12 & 1.12 & - \\
         \hline
         $\Xi_{bc}^{\prime+}\rightarrow\Xi_{bc}^{\ast+}$& $0.17\pm0.030$ & $-0.36$ & -0.36 & 0.672  & 1.04 & 0.814 & - \\
         \hline
         $\Xi_{bc}^0\rightarrow\Xi_{bc}^{\ast0}$& $0.77\pm0.104$ & $1.02$ &  1.03 & -0.747 & -1.03 & -0.919 & - \\
         \hline
         $\Xi_{bc}^{\prime0}\rightarrow\Xi_{bc}^{\ast0}$& $0.18\pm0.032$ & $-0.36$  & -0.36 & 0.070 & -0.17 & 0.598 & - \\
         \hline
         $\Xi_{cc}^{++}\rightarrow\Xi_{cc}^{\ast++}$& $1.03\pm0.138$ & $-1.4$  & -2.35 & -0.787 & -1.30 & -1.21 & -0.772\\
         \hline
         $\Xi_{cc}^+\rightarrow\Xi_{cc}^{\ast+}$& $0.96\pm0.158$ & $1.23$  & 1.55 & 0.945 & 1.19 & 1.07 & 0.906\\
         \hline
         $\Xi_{bb}^0\rightarrow\Xi_{bb}^{\ast0}$& $1.78\pm0.300$ & $-1.82$ & -2.77 & -1.039 & -1.70 & -1.045 & - \\
         \hline
         $\Xi_{bb}^-\rightarrow\Xi_{bb}^{\ast-}$& $0.82\pm0.131$ & $0.81$ & 1.13 & 0.428 & 0.76 & 0.643 & - \\
        \hline
    \end{tabular}
    \caption{Transition magnetic moments in nuclear magneton}
    \label{magnetic_results}
\end{table}
\begin{table}[ht!]
    \centering
    \begin{tabular}{|c|c|}
        \hline
         Transition & $|G_E|$  \\
         \hline
         $\Xi_{bc}^+\rightarrow\Xi_{bc}^{\ast+}$& $\left( 1.92\pm0.239\right)\times10^{-3}$\\
         \hline
         $\Xi_{bc}^{\prime+}\rightarrow\Xi_{bc}^{\ast+}$& $\left( 0.167\pm0.029\right)\times10^{-3}$\\
         \hline
         $\Xi_{bc}^0\rightarrow\Xi_{bc}^{\ast0}$& $\left( 1.18\pm0.160\right)\times10^{-3}$\\
         \hline
         $\Xi_{bc}^{\prime0}\rightarrow\Xi_{bc}^{\ast0}$& $\left( 0.18\pm0.032\right)\times10^{-3}$\\
         \hline
         $\Xi_{cc}^{++}\rightarrow\Xi_{cc}^{\ast++}$& $\left( 5.81\pm0.775\right)\times10^{-3}$\\
         \hline
         $\Xi_{cc}^+\rightarrow\Xi_{cc}^{\ast+}$& $\left( 5.44\pm0.901\right)\times10^{-3}$\\
         \hline
         $\Xi_{bb}^0\rightarrow\Xi_{bb}^{\ast0}$& $\left( 1.56\pm0.264\right)\times10^{-3}$ \\
         \hline
         $\Xi_{bb}^-\rightarrow\Xi_{bb}^{\ast-}$& $\left( 0.72\pm0.115\right)\times10^{-3}$ \\
        \hline
    \end{tabular}
    \caption{Transition electric quadrupole moments in nuclear magneton}
    \label{electric_results}
\end{table}
 All errors coming from uncertainties of input parameters, as well as the variation of $s_0$, $M^2$, and $\cos\theta$ in their working regions, are taken quadratically. For comparison in Table-\ref{magnetic_results}, we present the predictions for the magnetic moments $G_M(0)$ of other approaches such as quark model \cite{transition15}, $\chi$PT \cite{transition15}, bag model \cite{transition10}, effective quark mass scheme \cite{DhirEMS}, extended bag model \cite{exBag} and lattice QCD \cite{lattice}
Using the results on the moments $G_M$ and $G_E$ one can easily estimate the decay widths with the help of the following formula
\begin{equation}
    \Gamma = \dfrac{\alpha}{16}\dfrac{\left(m_1^2-m_2^2\right)^3}{m_N^2m_1^3}\left[G_M^2(0)+3G_E^2(0)\right]
    \label{decay_width_formula}
\end{equation}
where $m_N$ is the mass of the nucleon.
The results of the decay widths are collected in table-\ref{decays_table}.
\begin{table}[ht]
    \centering
    \begin{tabular}{|c|c|c|c|c|c|c|}
        \hline
         Decay & This Work & $\chi$ PT\cite{transition15} & Bag Model\cite{transition10} & exBag Model\cite{exBag} & RQM\cite{RQM} & Lattice\cite{lattice}  \\
         \hline
         $\Xi_{bc}^{\ast+}\rightarrow\Xi_{bc}^{+}\gamma$& $0.48\pm0.119$ & $26.2$ & 0.533 & 1.31 & - & -\\
         \hline
         $\Xi_{bc}^{\ast+}\rightarrow\Xi_{bc}^{\prime+}\gamma$& $0.002\pm0.0007$ & $0.52$ & 0.031 & 0.0293 & - & -\\
         \hline
         $\Xi_{bc}^{\ast0}\rightarrow\Xi_{bc}^{0}\gamma$& $0.18\pm0.049$ & $7.19$ &  0.612 & 0.876 & - & -\\
         \hline
         $\Xi_{bc}^{\ast0}\rightarrow\Xi_{bc}^{\prime0}\gamma$& $0.003\pm0.0009$ & $0.52$  & 0.000 & $7.6 \cdot 10^{-5}$ & - & -\\
         \hline
          $\Xi_{cc}^{\ast++}\rightarrow\Xi_{cc}^{++}\gamma$& $2.36\pm0.622$ & 22.0 & 1.43 & 2.79 & 7.21 & 0.0518\\
         \hline
         $\Xi_{cc}^{\ast+}\rightarrow\Xi_{cc}^{+}\gamma$& $2.07\pm0.666$ & 9.57 & 2.08 & 2.17 & 3.90 & 0.0648\\
         \hline
         $\Xi_{bb}^{\ast0}\rightarrow\Xi_{bb}^{0}\gamma$& $0.58\pm0.188$ & $31.1$ & 0.126 & 0.137 & 0.98 & -\\
         \hline
         $\Xi_{bb}^{\ast-}\rightarrow\Xi_{bb}^{-}\gamma$& $0.12\pm0.038$ & $5.17$ & 0.022 & 0.0268 & 0.21 & -\\
        \hline
    \end{tabular}
    \caption{Decay widths in keV}
    \label{decays_table}
\end{table}

For completeness, we also presented the decay width results obtained from the bag model,exBag model, RQM, and $\chi$PT. We see that our results are drastically different from the results of both approaches except the results for $\Xi_{bc}^{\ast+}\longrightarrow\Xi_{bc}^+\gamma$ and $\Xi_{bc}^{\ast0}\longrightarrow\Xi_{bc}^0\gamma$ transition obtained from bag model which are in agreement with our results. 

From Table-\ref{magnetic_results} we get the following results: Our finding of transition magnetic moments for $\Xi_{bc}^{\ast+}\rightarrow\Xi_{bc}^{+}$, $\Xi_{cc}^{++}\rightarrow\Xi_{cc}^{\ast++}$, $\Xi_{cc}^{+}\rightarrow\Xi_{cc}^{\ast+}$ and $\Xi_{bb}^{\ast-}\rightarrow\Xi_{bb}^{-}$ transitions are in agreement with the QM\cite{transition15}, EMS \cite{DhirEMS} and exBag \cite{exBag} but considerably differ from $\chi$PT\cite{transition15}. Results for the $\Xi_{bc}^{\ast0}\rightarrow\Xi_{bc}^{0}$ transition is in agreement with $\chi$PT \cite{transition15} and exBag \cite{exBag}, $\Xi_{bc}^{\ast0}\rightarrow\Xi_{bc}^{\prime0}$ transition is agreeing with the EMS \cite{DhirEMS}, $\Xi_{bc}^{\ast0}\rightarrow\Xi_{bc}^{0}$ transition is in agreement with bag model \cite{transition10} and exBag \cite{exBag}, and  $\Xi_{bb}^{\ast0}\rightarrow\Xi_{bb}^{0}$ agrees with the both QM \cite{transition15} and EMS \cite{DhirEMS}. While the results for $\Xi_{cc}^{++}\rightarrow\Xi_{cc}^{\ast++}$ also agrees with the quark model \cite{transition15} and $\Xi_{cc}^{+}\rightarrow\Xi_{cc}^{\ast+}$ agrees with the lattice qcd \cite{lattice}, results for $\Xi_{bc}^{\ast+}\rightarrow\Xi_{bc}^{\prime+}$ transition differs from other approaches results.

From Table-\ref{decays_table} we get the following results: Our finding of the decay widths for the $\Xi_{bc}^{\ast+}\rightarrow\Xi_{bc}^{+}$ transition is in agreement with the bag model\cite{transition10}, $\Xi_{cc}^{\ast++}\rightarrow\Xi_{cc}^{++}$ and $\Xi_{cc}^{\ast+}\rightarrow\Xi_{cc}^{+}$ are in agreement with the exBag model\cite{exBag} while $\Xi_{cc}^{\ast+}\rightarrow\Xi_{cc}^{+}$ transition also agrees with the bag model\cite{transition10}. For other transitions, the results differ from other works. As we stated earlier, the values of the decay widths are sensitive to the mass difference of the initial and final baryons, hence the results differ drastically with the selection of the masses.

\section{Conclusion}

We calculated the magnetic dipole $G_M(0)$ and electric quadrupole $G_E(0)$ formfactors of spin-$\nicefrac{3}{2}$ doubly heavy baryons to spin-$\nicefrac{1}{2}$ doubly heavy baryon transitions within light cone sum rules method. Having the values of the magnetic dipole $G_M(q^2)$ and electric quadrupole $G_E(q^2)$ formfactors at $q^2=0$ we estimate the corresponding decay widths. Our findings on formfactors and decay widths compared with the predictions of other approaches.   
%
\newpage

\tikzexternaldisable
\bibliography{ref}

\newpage

\appendices
\section{Expressions for the Spectral Densities}
\label{appndxb}
Here, we present the expressions for the spectral densities in this study.
\subsection{Spectral Densities for Symmetric Currents}
\begin{align}   
        \nonumber\rho_1^S(Q,Q^\prime,q) &= \dfrac{3(\beta-1)e_q}{32\sqrt{3}\pi^4} \int_0^1 dx \theta(s-t_{QQ^\prime}(x,\overline{x}))(s-t_{QQ^\prime}(x,\overline{x}))^2x\xbar\\
        \nonumber& +\int_0^1 dx \int_0^{1-x} dy \dfrac{3\theta\left(s-t_{QQ^\prime}(x,y)\right)}{64\pi^4}\Bigg\{\\
        \nonumber&(\beta-1)(s-t_{QQ^\prime}(x,y))^2\left[xy\bar{u}_0(2e_q - (e_{Q^\prime} +e_Q) )+(e_{Q^\prime}x+e_Qy)\right]\\
        \nonumber&-\dfrac{4m_Qm_{Q^\prime}\beta(s-t_{QQ^\prime}(x,y))}{3xy}\left(2e_qxy-(1-x-y)(e_{Q^\prime}x+e_Qy)\right)\\
        \nonumber&+\dfrac{2(\beta-1)\bar{u}_0(1-x-y)(s-t_{QQ^\prime}(x,y))}{xy}\left(e_{Q^\prime}m_{Q^\prime}^2x^2+e_Qm_Q^2y^2\right)\Bigg\}\\
        \nonumber&+ \dfrac{e_q}{48\sqrt{3}\pi^2} \int_0^1 dx \Bigg\{ 6\qbarq \bar{u}_0 (\beta+1) \int_{u_0}^1 du  h_\gamma(u)\rhonp{1,1} \\ 
        \nonumber&- 2f_1(\mathcal{A}) \dfrac{f_{3\gamma} (\beta-1)(s-t_{QQ^\prime}(x,\overline{x}))}{(m_Q+m_{Q^\prime})}\rhonp{0,0}\\
        \nonumber&\dfrac{(1+3\beta)\qbarq}{48\sqrt{3}\pi^2}\left[2f_2(\mathcal{T}_4^\gamma)\int_0^1dx\theta(s-t_{QQ^\prime}(x,\xbar))(e_{Q^\prime}m_Q+e_Qm_{Q^\prime})\right.\\
        &\left.-f_3(\mathcal{T}_4^\gamma)\int_0^1dx\theta(s-t_{QQ^\prime}(x,\xbar))\dfrac{e_{Q^\prime}\xbar+e_Qx}{x\xbar}(m_{Q^\prime}\xbar+m_Qx)\right]\\
        \nonumber&+\dfrac{\qbarq}{48\sqrt{3}\pi^2}f_3(\mathcal{S}^\gamma)\int_0^1\theta(s-t_{QQ^\prime}(x,\xbar)) \left[2\beta(e_Qm_{Q^\prime}+e_{Q^\prime}m_Q)\right.\\
        \nonumber&\left.-\dfrac{(1+\beta)(e_Qm_Qx^2+e_{Q^\prime}m_{Q^\prime}\xbar^2}{x\xbar}\right]\\
        \nonumber&+\qbarq\Bigg[ \Big((\beta+1) f_2\left(4\widetilde{\mathcal{S}}+2\mathcal{T}_2+12\mathcal{T}_3-12\mathcal{T}_4\right) - 4f_2\left(\widetilde{\mathcal{S}}+\nonumber\mathcal{T}_3-\mathcal{T}_4\right) \\
        &- (\beta+1)f_3\left(3\mathcal{T}_4 + 3\mathcal{S}+4\nonumber\mathcal{T}_1+5\mathcal{T}_2+\widetilde{\mathcal{S}}\right)\Big)\rhonp{0,0}\\
        &+\Big(2f_3\left(2\mathcal{T}_1+\mathcal{T}_3\right)-\nonumber(\beta+1)f_3\left(3\mathcal{T}_3+\widetilde{\mathcal{S}}+\mathcal{T}_4-\mathcal{S}-\mathcal{T}_1-\mathcal{T}_2\right)\Big)\rhonp{1,0}\Bigg]\\
        \nonumber&+\dfrac{3\qbarq(1+3\beta)}{2}\left(4\chi\varphi_\gamma(u_0)(s-t_{QQ^\prime}(x,\overline{x}))-(2+t_{QQ^\prime}(x,\overline{x}))\nonumber\mathbb{A}(u_0)\right)\rhonp{1,1}\\
        \nonumber&+\dfrac{f_{3\gamma}}{(m_Q+m_{Q^\prime})}\left[6\psi^a(u_0)\left((\beta-1)(3s-2t_{QQ^\prime}(x,\overline{x}))\rhonp{0,1}+2\beta m_Qm_{Q^\prime}\rhonp{0,0}\right)\right.\\
        \nonumber&\left.+3(\beta-1)\left(4\psi^v(u_0)-\psi^{a\prime}(u_0)\right)(s-t_{QQ^\prime}(x,\overline{x}))\bar{u}_0\rhonp{0,1}\right]\Bigg\}    
\end{align}
\begin{align}
        \nonumber\rho_2^S(Q,Q^\prime,q) &= \int_0^1 dx \int_0^{1-x} dy \dfrac{3(1-\beta)u_0\bar{u}_0\theta\left(s-t_{QQ^\prime}(x,y)\right)}{16\sqrt{3}\pi^4}\\
        \nonumber&\Bigg\{(s-t_{QQ^\prime}(x,y))\Big[(1-x-y)(e_{Q^\prime}x+e_Qy) - 2xye_q\Big]\Bigg\}\\
        \nonumber&+\dfrac{\qbarq(1-\beta)}{12\sqrt{3}\pi^2}f_4(\mathcal{T}_4^\gamma)\int_0^1dx\delta(s-t_{QQ^\prime}(x,\xbar)\dfrac{e_{Q^\prime}m_{Q^\prime}\xbar^2+e_Qm_Qx^2}{x\xbar}\\
        \nonumber&+\dfrac{e_q}{24\sqrt{3}} \int_0^1 dx\Bigg\{\\ 
        &\dfrac{\theta\left(s-t_{QQ^\prime}(x,\overline{x})\right)(\beta-1)f_{3\gamma}}{(m_Q+m_{Q^\prime})}\Bigg[12\int_{u_0}^1 du  \bar{u} \psi^v(u) \rhonp{0,1}\\
        \nonumber&+3 u_0 \bar{u}_0 \psi^a(u_0) \rhonp{0,1} -(f_3-f_2)(\mathcal{A}+\mathcal{V})\rhonp{0,0}\Bigg]\\
        \nonumber&+\delta\left(s-t_{QQ^\prime}(x,\overline{x})\right)\qbarq\Bigg[6(1+\beta)\int_{u_0}^1 du \bar{u} (u-u_0) h_\gamma(u) \rhonp{1,1}\\
        \nonumber&+2f_4(\mathcal{T}_4-\mathcal{T}_2) (\beta+1)  \rhonp{0,0}+2f_4(\mathcal{T}_1-\mathcal{T}_3) (\beta-1) \rhonp{2,1}\Bigg]\Bigg\}
    \end{align}
\subsection{Spectral Densities for Antisymmetric Currents}
\begin{align}
        \nonumber\rho_1^A(Q,Q^\prime,q) &= \dfrac{1}{32\pi^4}\int_0^1 dx \int_0^{1-x} dy \theta(s-t_{QQ^\prime}(x,y))(s-t_{QQ^\prime}(x,y))\Bigg\{\\
        \nonumber&\dfrac{1-x-y}{xy}\Big[3(\beta-1)\bar{u}_0(e_Qm_Q^2y^2-e_{Q^\prime}m_{Q^\prime}^2x^2)-(4\beta+2)m_Qm_{Q^\prime}(e_Qy-e_{Q^\prime}x)\Big]\\
        \nonumber&-3(\beta-1)(s-t_{QQ^\prime}(x,y))\Big[xy\bar{u}_0(e_Q-e_{Q^\prime})-(e_Qy-e_{Q^\prime}x)\Big]\Bigg\}\\
        \nonumber&+\dfrac{(1-\beta)\qbarq}{144\pi^2}\left[2f_2(\mathcal{T}_4^\gamma)\int_0^1dx\theta(s-t_{QQ^\prime}(x,\xbar))(e_{Q^\prime}m_Q-e_Qm_{Q^\prime})\right.\\
        \nonumber&\left.-f_3(\mathcal{T}_4^\gamma)\int_0^1dx\theta(s-t_{QQ^\prime}(x,\xbar))\dfrac{e_{Q^\prime}\xbar+e_Qx}{x\xbar}(m_Qx-m_{Q^\prime}\xbar)\right]\\
        \nonumber&+\dfrac{\qbarq}{144\pi^2}f_3(\mathcal{S}^\gamma)\int_0^1\theta(s-t_{QQ^\prime}(x,\xbar)) \left[6\beta(e_Qm_{Q^\prime}-e_{Q^\prime}m_Q)\right.\\
        &\left.-\dfrac{(1+5\beta)(e_Qm_Qx^2-e_{Q^\prime}m_{Q^\prime}\xbar^2}{x\xbar}\right]\\
        \nonumber&+\dfrac{1}{144\pi^2}\int_0^1 dx \Bigg\{\dfrac{2f_1(\mathcal{V}) f_{3\gamma} (\beta-1)(s-t_{QQ^\prime}(x,\overline{x}))}{(m_Q-m_{Q^\prime})}\rhonm{0,0}\\
        \nonumber&+\qbarq\Bigg[2\Big( (\beta-1)f_3(\mathcal{T}_4-\mathcal{T}_3-2\widetilde{\mathcal{S}}+\mathcal{T}_2)+6f_3(\mathcal{T}_2-\widetilde{\mathcal{S}}) \Big)\rhonm{0,0}\\
        \nonumber&+\Big( (\beta-1)f_3(\mathcal{T}_2+\mathcal{T}_4+\mathcal{S}-3\widetilde{\mathcal{S}}) + 6f_3(\mathcal{T}_2-\widetilde{\mathcal{S}})\Big)\rhonm{2,1}\\
        \nonumber&+\Big( (\beta-1)f_3(\mathcal{T}_3-\mathcal{T}_1+4\mathcal{T}_4-6\mathcal{S}+2\widetilde{\mathcal{S}}+6\mathcal{T}_2)+6f_3(\mathcal{T}_4-\mathcal{S}+\widetilde{\mathcal{S}}-\beta\mathcal{T}_2)\Big)\rhonm{1,0}\\
        \nonumber&+\dfrac{3(1-\beta)}{2}\left(4(s-t_{QQ^\prime}(x,\overline{x}))\chi \varphi_\gamma(u_0) - (2+t_{QQ^\prime}(x,\overline{x}))\mathbb{A}(u_0)\right)\rhonm{1,1}\\
        \nonumber&+6\bar{u}_0(\beta+5) \int_{u_0}^1 du  h_\gamma(u) \rhonm{1,1}\Bigg]\Bigg\}
\end{align}
\begin{align}
        \nonumber\rho^A_2(Q,Q^\prime,q) &= \int_0^1 dx \int_0^{1-x} dy \\
        \nonumber&\dfrac{3(\beta-1) u_0\bar{u}_0\delta\left(s-t_{QQ^\prime}(x,y)\right)}{16\pi^4}\Big[(s-t_{QQ^\prime}(x,y)) (1-x-y)(e_{Q^\prime} x - e_Q y) \Big]  \\
        \nonumber&+\dfrac{\qbarq(1-\beta)}{36\pi^2}f_4(\mathcal{T}_4^\gamma)\int_0^1dx\delta(s-t_{QQ^\prime}(x,\xbar)\dfrac{e_{Q^\prime}m_{Q^\prime}\xbar^2-e_Qm_Qx^2}{x\xbar}\\
        \nonumber&\dfrac{e_q}{36\pi^2} \int_0^1 dx\Bigg\{  \\
        &\dfrac{3(f_2+f_3)(\mathcal{V}+\mathcal{A}) f_{3\gamma} (1-\beta) }{2\pi^2(m_Q+m_{Q^\prime})}\theta\left(s-t_{QQ^\prime}(x,\overline{x})\right) (x-\xbar)\rhonm{0,0}\\
        \nonumber&+\delta\left(s-t_{QQ^\prime}(x,\overline{x})\right)e_q\qbarq\Bigg[3(\beta+5)\int_0^1 du \bar{u}_0(u-u_0)\theta(u-u_0)h_\gamma(u) \rhonm{1,1}  \\
        \nonumber&-f_4((1-vthe)(\mathcal{T}_4-\mathcal{T}_2))  (\beta+5) \rhonm{0,0}+\dfrac{3f_4(\mathcal{T}_1-\mathcal{T}_3) (\beta-1)}{2} \rhonm{2,1}  \\
        \nonumber&+f_4(\mathcal{T}_4-\mathcal{T}_2) \Big[ (\beta+5)\rhonm{2,1} - 3(\beta+1)\rhonm{1,0} \Big]\Bigg]\Bigg\} 
    \end{align}
\begin{equation}
    \rho^{(\pm)}_{n,m} =  \theta\left(s-t_{QQ^\prime}(x,\overline{x})\right) (x\xbar)^m \left(\dfrac{m_Q}{\xbar^n}\pm\dfrac{m_{Q^\prime}}{x^n}\right) 
\end{equation}

\begin{equation}
    \begin{aligned}
        f_1(\varphi) &= \int d\ali\frac{\varphi(\alpha_i)}{\alg^2}\theta\left(u_0-\alqb\right)\left[\theta\left(\alg+\alqb-u_0\right)-\alg\theta\left(\alg\right)\delta\left(\alg+\alqb-u)\right) \right] \\
        f_2(\varphi) &= \int d\ali\frac{(\alg+\alqb-u_0) \varphi(\alpha_i)\theta(\alg+\alqb-u_0)\theta(u_0-\alqb)}{\alg^2} \\
        f_3(\varphi) &=\int d\ali\frac{\varphi(\alpha_i)\theta(\alg+\alqb-u_0)\theta(u_0-\alqb)}{\alg} \\
        f_4(\varphi) &= \int d\ali\int_0^1 dv \varphi(\alpha_i) \theta(-v\alg+\alg+\alqb-u_0)) \\
    \end{aligned}
\end{equation}

where $t_{QQ^\prime}(x,\overline{x}) = \dfrac{m_Q^2}{\xbar}+\dfrac{m_{Q^\prime}^2}{x}$,  $t_{QQ^\prime}(x,y) = \dfrac{m_Q^2}{x}+\dfrac{m_{Q^\prime}^2}{y}$

\end{document}